\documentclass[12pt]{article}
\usepackage{arxiv}

\usepackage[utf8]{inputenc} 
\usepackage[T1]{fontenc}    
\usepackage{hyperref}       
\usepackage{url}            
\usepackage{booktabs, threeparttable}       
\usepackage{amsfonts, amssymb, amsmath, latexsym, amsthm, bm}       
\usepackage{nicefrac}       
\usepackage{microtype}      
\usepackage{lipsum}		
\usepackage{graphicx}
\usepackage{natbib}
\usepackage{doi}
\usepackage{pdflscape}
\usepackage{listings}
\title{A penalized heteroskedastic ordered probit model for DIF (measurement invariance) testing of single-item assessments in cross-cultural research}
\author{ \href{https://orcid.org/0000-0002-9114-3896}{\includegraphics[scale=0.06]{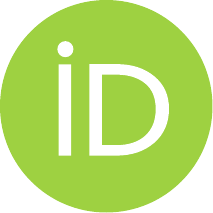}\hspace{1mm}R. Noah Padgett} \\
	Department of Epidemiology\\
	Harvard T.H. Chan School of Public Health\\
        Harvard University\\
	Boston, MA  \\
	\texttt{npadgett@hsph.harvard.edu}\\
}

\renewcommand{\shorttitle}{Penalized HETOP for DIF}

\hypersetup{
pdftitle={A penalized heteroskedastic ordered probit model for DIF/MI of single-item assessments in cross-cultural research},
pdfsubject={stat.ME},
pdfauthor={R. Noah Padgett},
pdfkeywords={Bayesian, SEM, factor analysis, item response theory, differential item functioning, regularization, penalty, heteroskedastic, ordered probit},
}

\begin{document}
\large
\maketitle

\begin{abstract}
Differential item functioning (DIF) or measurement invariance (MI) testing for single-item assessments has previously been impossible. Part of the issue is that there are no conditioning variables to serve as a proxy for the latent variable--regression-based DIF methods. Another reason is that factor-analytic approaches require multiple items to estimate parameters. In this technical working paper, I propose an approach for evaluating DIF/MI in a single-item assessment of a construct. The current methods should NOT replace using multiple-indicator MG-CFA/IRT analyses of DIF/MI or regression mased methods when possible. More items generally provide significantly better construct coverage and provide more rigorous DIF/MI evaluation.
\end{abstract}


\section*{Introduction}

A foundational component of nearly all scale development processes is investigating measurement invariance across subpopulations. 
Invariance of scale properties is sometimes referred to as fairness in an educational setting, as defined by the Standards (AERA, APA, NCME, 2024, p. 50) or to provide validity evidence for the consistency in interpretation of scores across groups of like respondents.
Many veins of research have developed to help empirically assess the degree of validity evidence in support (or against) the conclusion that scores are comparable across groups \citep{Van_Erp2018, van-de-Vijver2019-ar}.

One way validity evidence can be provided is by evaluating the equivalence of sets of parameters in a confirmatory factor model across two or more groups (e.g., is the entire factor loading matrix equal for males and females? Or, are all item residual variances equal over time?).
Researchers have provided some excellent guidance and recommendations for scale developers to use, building on this perspective \citep{Vandenberg2000-vo}.
However, these methods can be clunky when trying to identify partial invariance, which is the likely scenario where only a subset of parameters vary meaningfully across groups.

Measurement model parameters varying across groups is called measurement non-invariance.
This means that the characteristics of the models theorized to govern the differences in why people have different responses to the same item differ according to a known characteristic.
Said another way, the moments of the observed data (e.g., means, variances, covariances) differ as a function of the statistical measurement model AND group membership. 
\citet{Millsap2012} describes this as the expectation ($\mathbb{E}(.)$) and variance-covariance ($\mathbb{V}(.)$) are only conditional on the set of parameters ($\theta$) governing the measurement model and not conditional on a grouping variable ($G$):
\begin{align}
    \mathbb{E}\left[Y \vert \theta, G\right] &= \mathbb{E}\left[Y\vert \theta \right]\\
    \mathbb{V}\left[Y \vert \theta, G\right] &= \mathbb{V}\left[Y\vert \theta \right]
\end{align}
A difficulty in determining the extent to which the above may be violated is that we must estimate the parameters ($\theta$), which are typically not known a priori.
Our question for single-item assessments is, therefore, what parameters ($\theta$) can we estimate meaningfully?
Next, we explore how the statistical model for measurement based on factor analysis can be expressed when only a single-observed variable is observed.

\section*{Single-item measurement models}

This paper focuses on single-item measures with any number of discrete categories or levels. The more categories, the more precision we have in identifying the specific location of each group on the construct of interest. We will build up the statistical measurement model in stages starting with classical test theory.
The classical test theory equation relates the observed scores on the test ($Y$) to the observed true score on the test ($T$) with an error component ($E$): 
\begin{align}
Y &= T + E
\end{align}
This forms the basis for the extension to a factor analytic representation where the true score ($T$) is replaced by a latent variable ($\eta$) that is not necessarily a perfectly reflected by the observed variable measure as a function of error ($\varepsilon$) and reliability ($\lambda$).
The latent variable $\eta$ is commonly assumed to follow some distribution in the population, such as $\eta\sim\text{Normal}(\alpha,\phi)$ with some location and variability.
Combined, these pieces lead to the conceptual model for a single observed variable $Y$ being
\begin{align}
    Y &= \lambda\eta + \varepsilon.
\end{align}
The statistical component of the model assumes that some parameters follow a distribution that characterizes the sources of variability in the observed scores
\begin{align}
    \eta &\sim \text{Normal}(\alpha, \phi)\\
    \varepsilon &\sim \text{Normal}(0, \sigma).
\end{align}
These pieces can then further be combined to the implied distribution of the observed scores, which can be conditional on the value of the latent variable
\begin{align}
    Y \vert \eta &\sim \text{Normal}\left(\lambda\eta, \sigma\right).
\end{align}
or marginalized over the latent variable
\begin{align}
    Y &\sim \text{Normal}\left(\lambda\alpha, \lambda^2\phi + \sigma\right).]\label{eq:marginal}
\end{align}
Unfortunately, the latent variable $\eta$ cannot be observed directly, resulting in the above model not being statistically identifiable with unique parameter values for all the components ($\sigma$-item residual variance, $\lambda$-item loading or item reliability, $\alpha$-latent variable location, and $\phi$-latent variable variance).
In multi-item measurement models, several well-known methods exist for identifying the model under different conditions \citep{Bollen1989}.
The identification constraints can generally be simplified to the rule of thumb that the number of scale/variance parameters you can identify is equal to the number of unique elements of the variance-covariance matrix of the set of observed variables and the number of location parameters you can identify is equal to the number of variables you have (if continuous) or the number of categories minus 1 if categorical.
However, for single-item assessments, these conditions do not apply, and we have only the moments of a single observed variable to manipulate.

Single-item assessments have
\begin{itemize}
    \item a location ($\mu$) that is a function of item reliability ($\lambda$) and the latent variable mean ($\alpha$), and
    \item a scale that is a function of item reliability ($\lambda$), factor variance ($\phi$), and item residual variance ($\sigma$).
\end{itemize}
Resulting in a total of four distinct measurement model parameters that could reasonably vary across groups.
However, an observed variable's distribution can be completely characterized using a location parameter (sample mean) and a scale parameter (sample standard deviation).
With a single observed variable, there is no way to form a sample estimator for item reliability to the best of our knowledge. 

Suggested simplifications is to 
\begin{itemize}
    \item For the item location, the group specific locations are centered at the item overall average such that deviations around the grand mean average to zero, a "centered" parameterization. 
    \item We cannot, to the best of my knowledge, differentiate the factor loading and latent response total variance with only a single item. We therefore have to decide how to parameterize the model for interpretation. One way that may be especially useful for our purposes is an IRT-like parameterization of 2-parameter models for binary outcomes. From equation \ref{eq:marginal}, the probability of given range of Y values is
    \begin{align*}
        Pr(Y < \tau) = \Phi\left(\frac{\lambda\alpha - \tau}{\lambda^2\phi + \sigma}\right)
    \end{align*}
    However, if we assume that the total variance is 1.0, the model simplifies to 
    \begin{align*}
        Pr(Y < \tau) = \Phi\left(\lambda\alpha - \tau\right)
    \end{align*}
    which is similar in style of the traditional 2-parameter normal-ogive model from item response theory \citep{Kamata2008}. The traditional model is simply $Pr(Y=1|\theta)=\Phi[\alpha(\theta - \beta)]$, so there is a similarity we can use to help motivate the use of this parameterization if we are willing to assume a fixed value for the variance. 
    Alternatively, we can evaluate DIF with respect to the total variance but not the item discrimination. 
\end{itemize}

\section*{Heteroskedastic Ordered Probit Model}

The heteroskedastic ordered probit model is a generalization of the traditional probit model to variables with more than two ordered categories.
The model is ``heteroskedastic'' in that the model definition allows for the scale in addition to the location of the underlying normally distributed ``latent response variable'' to vary according to some set of predictors.
The explanatory variables that inform location and scale do not necessarily need to be the same.
In practice, when different explanatory variables are used to inform the different parameters, it is better for identification.

The model for the probability of responding $k$ or lower (k=0,1,2,...,K) on variable $Y$ as a function of explanatory variables $\mathbf{x}$ and $\mathbf{z}$, where $\mathbf{x}$ and $\mathbf{z}$ can overlap or be distinct, leads to,
\begin{align}
    \text{Pr}\left(Y = k\right) &= \text{Pr}\left(Y^\ast < \tau_{k}\right) - \text{Pr}\left(Y^\ast < \tau_{k+1}\right)\\
    &=\Phi\left\lbrace\frac{\mathbf{x}\bm\beta - \tau_{k}}{\exp(\mathbf{z}\bm\gamma)}\right\rbrace - \Phi\left\lbrace\frac{\mathbf{x}\bm\beta - \tau_{k+1} }{\exp(\mathbf{z}\bm\gamma)}\right\rbrace\\
    Y = k &\Leftrightarrow \tau_{k} < Y^\ast \leq \tau_{k+1}\\
     Y^\ast &\sim \text{Normal}\left(\mathbf{x}\bm\beta, \exp(\mathbf{z}\bm\gamma)\right)
\end{align}
and $\Phi(.)$ is the cumulative distribution function of the standard normal distribution. The threshold parameters, $\tau_{k+1}$, for $k=0,1,...,K$, define the "location" on average dividing the propensity of responding to category k or lower verses k+1 or higher. Note that $\tau_0=-\infty$ and $\tau_{K+1}=\infty$.

The model is described quite well in the Stata documentation \citep{hetop-stata} and by \citet{Reardon2017-xb}.4
This model is similar in flavor to the graded response model \citep{Samejima1969-fj}, but developed separately to the best of my knowledge. I came across this heteroskedastic ordered probit model in an article by \citet{Reardon2017-xb} for using the model with coarsened educational attainment data for studying group ordering \citep{Padgett2025-op}.

The log-likelihood function is
\begin{equation}
    \ln L=\sum_{n=1}^{N}w_n \sum_{k=0}^{K} I(y_n=k)\ln \left[\Phi\left\lbrace\frac{\mathbf{x}_n\bm\beta - \tau_{k}}{\exp(\mathbf{z}_n\bm\gamma)}\right\rbrace - \Phi\left\lbrace\frac{ \mathbf{x}_n\bm\beta - \tau_{k+1}}{\exp(\mathbf{z}_n\bm\gamma)}\right\rbrace\right]
\end{equation}
This log-likelihood function includes the case-weight $w_n$.

The above parameterization is excellent if you have case-level data and when the sampling design is consistent across cases.
However, this is not always the case and sometimes you only have assess to summary statistics of group-level counts for each cell of the ``design matrix'' implied by $\mathbf{x}$ and $\mathbf{z}$.
This is where the develops by \citet{Reardon2017-xb} are incredibly useful.
They showed how, conditional on the group $g$ implied by the design matrix, the log-likelihood can be written in terms of cell sizes, provided no cells are empty:
\begin{align}
    \ln L &= \sum_{g=1}^Gw_g\left\lbrace\ln\left(n_g!\right) + \sum_{k=0}^{K}\left[n_{gk}\ln\left(\text{Pr}(y_g=k)\right)-\ln\left(n_{gk}!\right)\right]\right\rbrace\\
    &= \sum_{g=1}^Gw_g\sum_{k=0}^{K}n_{gk}\ln\left[\Phi\left(\frac{ \mu^\ast_g - \tau_{k}}{\sigma^\ast_g}\right) - \Phi\left(\frac{\mu^\ast_g - \tau_{k+1}}{\sigma^\ast_g}\right)\right] + C
\end{align}
where the last term is a constant $C=\ln\left(\frac{\prod_{g=1}^{G}n_g!}{\prod_{g=1}^{G}\prod_{k=0}^{K}n_{gk}!}\right)$. 
There is a subtle shift that happens when the model is parameterized in terms of the group-level summary statistics where the person-level characteristics $\mathbf{x}$ and $\mathbf{z}$ are replaced by the number of people or units of analysis in the cell of the joint design matrix implied by $\mathbf{x}$ and $\mathbf{z}$.
The location of group $g$ on the underlying latent response variable is $\mu^\ast_g = \mathbf{x}\bm\beta$.
The scale, or standard deviation of group $g$ on the underlying latent response variable is $\sigma^\ast_g=\exp(\mathbf{z}\bm\gamma)$.

Unfortunately, whether we use person-level data or group-summary statistics, we cannot recover $\mu^\ast_g$ and $\sigma^\ast_g$ without additional assumptions \citep[][, p. 8]{Reardon2017-xb}.
Several options are available to resolve the indeterminacy in location and scale parameters.
For location, we can set a ``reference'' group to be fixed at zero. Similar to how in Rasch modeling, one of the item difficulty parameters may be fixed to zero.
Another approach is to use a ``sum-to-zero'' approach that sets the average location to zero.
This is accomplished by freely estimating all but one group location and then fixing that group's location to the negative sum of all other groups.
For the scale, we can use similar approaches.
We can fix one ``reference'' group to have a referent standard deviation of 1.0, then all other groups' scale is relative to this focal group.
Another option is to use the ``sum-to-zero'' approach, but remember, here the deviations are computed on the log-scale to ensure the resulting scale estimates are strictly positive.

The model above allows us to estimate, on a common scale, the mean and standard deviation of the ``latent variable'' implied by a single item. In testing, I've found that the parameterization in (12) provides the fastest way to recover the moments while ensuring the full complex sampling design is accounted for.
A major advantage of this is that all aspects of the design (weights, strata, and PSUs) are incorporated by estimating the ``population count'' of each response category prior to using the HETOP model. 
Of particular benefit is that ``missing'' can be treated as a separate category in this step, side-stepping the need to explicitly model missing data, missing/nonresponse is treated as a population characteristic in the estimation.

\section*{Single-Item DIF Assessment}

The above model is helpful for recovering the implied location $\mu^\ast$ and variability $\sigma^\ast$ for each group $g$.
We could use those two recovered moments to evaluate the differences between groups after adding a penalty for those group differences.
However, for DIF assessment, evaluating the scale/variability of the latent response distribution variability is not the most frequently meaningful.
Instead, we can reparameterize the model in terms of the location and discrimination.

Discrimination is a common parameter for item response theory models of the probability of endorsing a response:
\begin{align*}
    Pr(Y=1\vert \theta) &= \Phi\left[\lambda(\theta - \beta)\right]
\end{align*}
where $\theta$ is the latent variable, $\beta$ is the item difficulty or location parameter, and $\lambda$ is the item discrimination or slope of the item characteristic curve (more commonly denoted as $\alpha$, but I used $\lambda$ here to illustrate the connection to the factor analytic tradition). 
The above model is a normal-ogive parameterization since it uses the cumulative normal distribution to define the probability of endorsing the item, but the logistic function is also a common choice.
When we only have a single item, the common estimation methods cannot be used to provide estimates for the parameters.
This is where the HETOP approach comes in to provide a path to evaluate group differences in these characteristics.
We first "marginalize" the latent variable out to only have a group summary statistic for the location: $\theta \rightarrow \mu^\ast$.
The item location is then only characterized by the "difficulty" thresholds: $\beta \rightarrow \tau_k$.
The item discrimination is then the inverse of the latent response variable standard deviation: $\lambda \rightarrow \frac{1}{\sigma^\ast}$.

When the HETOP model is reparameterized in this way, the likelihood can be rewritten as a function of the group locations $\bm\mu^\ast=[\mu^\ast_1, \mu^\ast_2, ..., \mu^\ast_G]$ and the group-specific discrimination parameters $\bm\lambda^\ast=[\lambda^\ast_1, \lambda^\ast_2, ..., \lambda^\ast_G]$
\begin{align}
    \ln L_{p} &= \sum_{g=1}^Gw_g\sum_{k=0}^{K}n_{gk}\ln\left[Pr(Y=k\vert\mu^\ast_g,\lambda^\ast_g)\right] + C + P_0(\bm\mu^\ast) + P_0(\bm\lambda^\ast)\\
    &Pr(Y=k\vert\mu^\ast_g,\lambda^\ast_g) = \Phi\left(\lambda^\ast_g(\mu^\ast_g - \tau_{k})\right) - \Phi\left(\lambda^\ast_g(\mu^\ast_g - \tau_{k+1})\right)\\
    P_0(\theta) &= \sum_{\forall g \in G}\sum_{\forall h \neq g \in G} \frac{f(\theta_g, \theta_h)}{\nu}\\
    f(\theta_g, \theta_h) &= {(\theta_g - \theta_h)}^2, \text{ridge penalty}\\
    f(\theta_g, \theta_h) &= \sqrt{{(\theta_g - \theta_h)}^2+\varepsilon}, \text{lasso penalty}\\
    f(\theta_g, \theta_h) &= \sqrt{\vert\theta_g - \theta_h\vert+\varepsilon)}, \text{alignment penalty}
\end{align}
The different penalty methods are discusses in \citet{asparouhov2024penalized} in the context of penalized structural equation models.
For any of the penalty methods, changing the magnitude of $\nu$ changes the degree of regularization. 
For out purposes, we will utilize the alignment approach, but more work is needed to explore how the choice of penalty function influence DIF assessment of single-item measures.
I recommend using many values of $\nu$ to identify how the estimates of $\bm\mu^\ast, \bm\lambda^\ast$ are sensitive to the regularization parameter $\nu$.

Another key piece of information to evaluate is the relative contribution of the penalty to the likelihood. 
\citet{asparouhov2024penalized} recommended that the penalty contribution should be no more than 5-10\% of the likelihood, but it's unclear why they recommended this threshold.
We recommend using the relative contribution, $\frac{P_0(\bm\mu^\ast) + P_0(\bm\lambda^\ast)}{\ln L_{p}}$, as a sensitivity test by plotting the magnitude $\nu$ and the proportion to see if there is spike or sharp change in the proportion.
Elbow(s) in the plot may point to the values of $\nu$ that balance the regularization with the data.

\subsection*{Determining DIF}

\begin{figure}[!htp]
    \centering
    \includegraphics[width=0.9\linewidth]{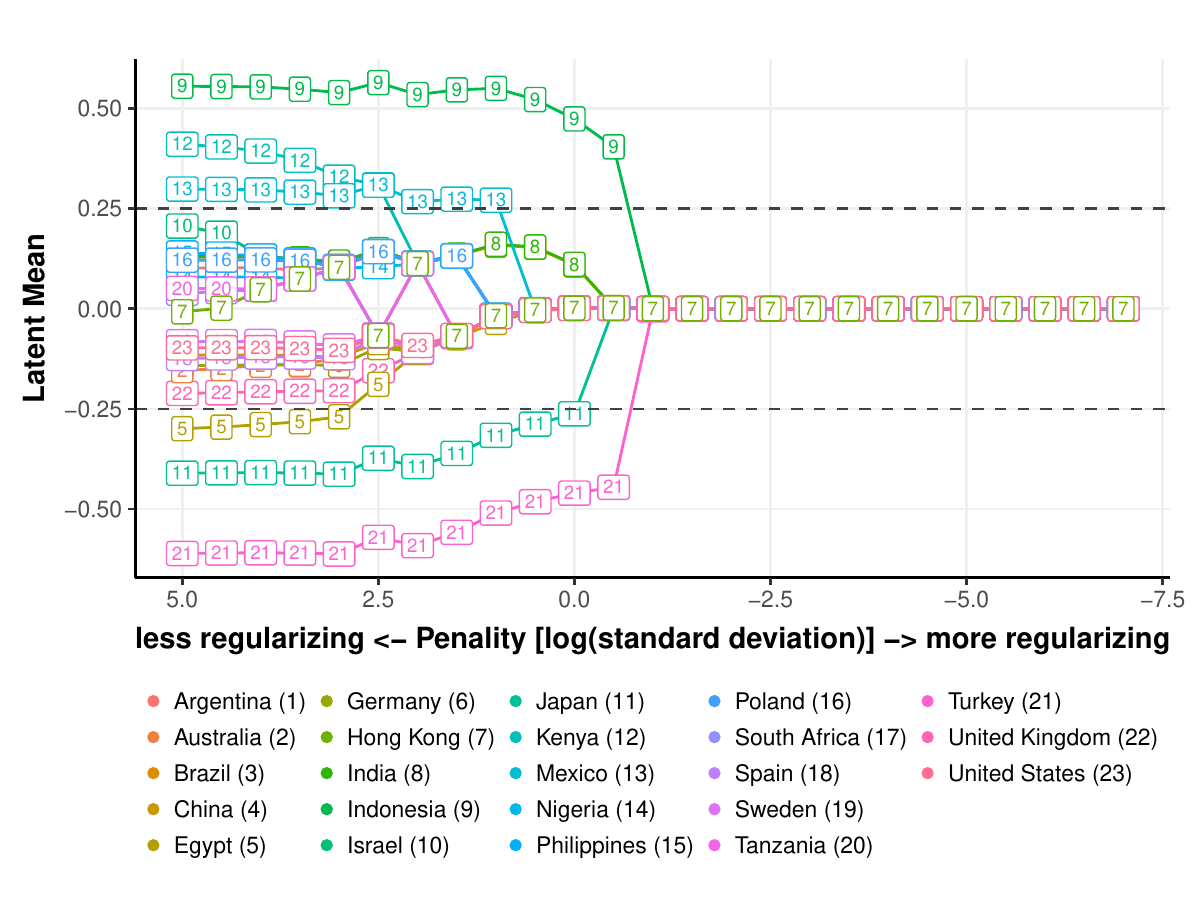}
    \caption{Item Happy--HETOP recovered latent means across countries by penalty with bounds at $\pm0.25$ for DIF across a range of penalty values.}
    \label{fig:happy-mu}
\end{figure}

DIF can be determined through a combination of examining the magnitude of the penalty and the magnitude of the difference between the baseline and group-specific estimates of $\mu_g^\ast$ and $\lambda^\ast_g$.
Figure \ref{fig:happy-mu} illustrates how the latent means for each group change as a function of the regularizing penalty parameter (log(standard deviation)), but doesn't provide guidance on identify which groups differ. 
To assist with this task, I recommend using a combination rule of
\begin{itemize}
    \item threshold of 10\% difference from baseline to determine a ``meaningful deviation'' from the population average of each parameter (mean or factor loading), and
    \item whether the $(1-\alpha/2)$\% CI for the estimate contains zero.
\end{itemize}
When used together, Figure \ref{fig:happy-mu} can be modified to include "bounds" to denote what we determine to be a sufficiently deviant from baseline to say, "wow, that's not close enough."
I recommend the following two bounds for the latent mean and discrimination parameters:
\begin{itemize}
    \item \textit{Latent mean}. The baseline latent mean is 0.0. So, I recommend a  "threshold" for sufficiently deviant that results in a 10\% change in the probability implied by the latent response variable. We solve the following for $x_0$:
    \begin{align}
        \vert\Phi(x_0) - \Phi(0)\vert &= 0.10\\
        x_0 &= \pm 0.255
    \end{align}
    \item \textit{Discrimination}. The baseline discrimination is 1.0. I recommend a ``threshold'' for sufficiently deviant that results in a 10\% change in the magnitude of the parameter value. So, any groups' $\lambda^\ast_g$ that is $>1.10$ or $<0.90$ would be a candidate for DIF in the discrimination.
\end{itemize}

Figure \ref{fig:happy-mu} illustrates this by plotting $\bm\mu^\ast$ across a range of values for $\log(\nu)$. 
We can see that, across a wide range of penalty values, countries Indonesia, Japan, Mexico, Japan, and Turkey each were outside the bounds of $\pm0.25$.
While Kenya and Egypt were outside the bounds when the penalty was relatively small.
this provides some evidence of sufficiently different latent means, or uniform DIF, for 4-6 countries depending on the strength on the penalty.
We could next select a specific value of penalty term ($\nu$) to estimate the model hessian and obtain approximate standard errors to evaluate more precisely which group(s) are significantly different than zero.

\begin{figure}[!htp]
    \centering
    \includegraphics[width=0.9\linewidth]{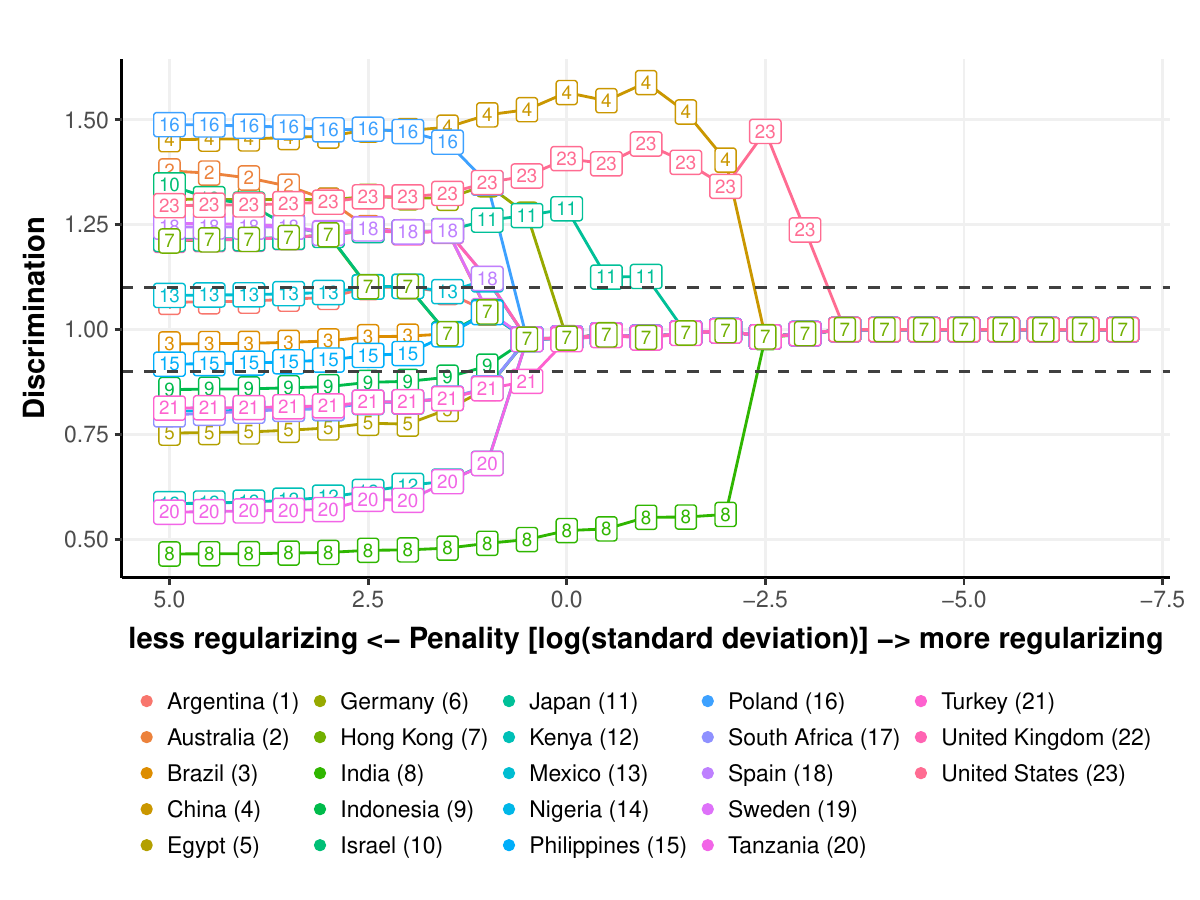}
    \caption{Item Happy--HETOP recovered item discrimination parameters across countries by penalty with bounds at $\pm10$\% for DIF across a range of penalty values.}
    \label{fig:happy-lambda}
\end{figure}

Figure \ref{fig:happy-lambda} illustrates how the item discrimination parameters vary across countries.

\subsection*{HETOP Item Characteristic Curves}
The item characteristic curve (ICC) are among the primary outputs from employing the HETOP model.

ICC, the resulting group means and discrimination parameters can be used to examine how, for people with the same level of the "latent" value of the construct, the expected item score changes.
The implied response is a straightforward sum of the probability of endorsing each category or lower for a given value of the latent variable $\theta$:
\begin{align}
    E[Y\vert \theta] &= \sum_{k=1}^{K-1} Pr(Y \leq k)\\
    &=\sum_{k=1}^{K-1} \Phi{\left[\lambda^\ast_g\left( \theta + \mu^\ast_g - \tau_{k}\right)\right]}
\end{align}

\begin{figure}[!htp]
    \centering
    \includegraphics[width=0.9\linewidth]{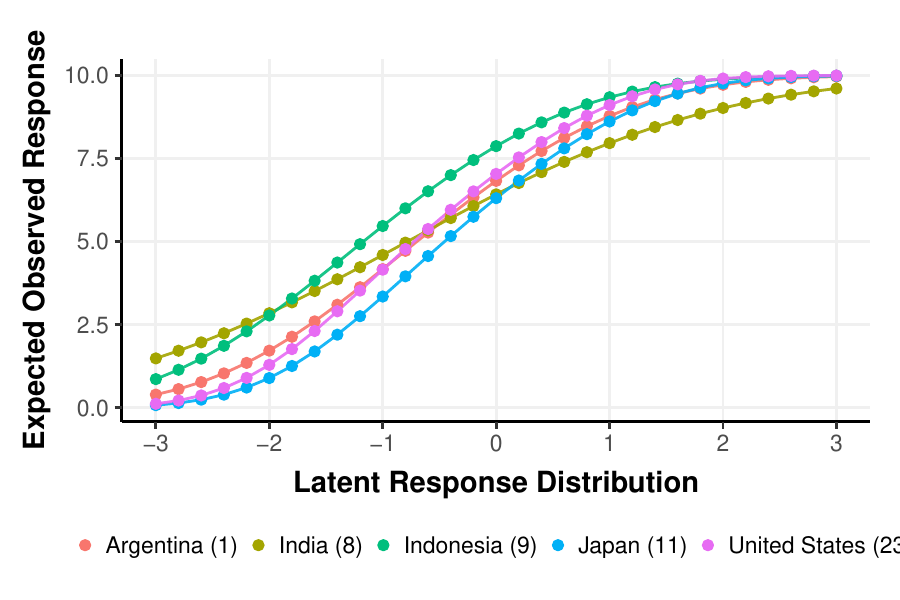}
    \caption{Item characteristic curves for Happy--recovered from the HETOP model at penalty $\nu=1.0$.}
    \label{fig:happy-icc}
\end{figure}

The resulting item characteristic curve can then be plotted for each group to see how moderated latent group means ($\mu^\ast_g$) and discrimination parameter ($\lambda^\ast_g$) change the shape of the item characteristic curve.
The item characteristic curve for the item Happiness is shown in Figure \ref{fig:happy-icc} for a subset of the countries with identified DIF.
I omitted China because the curve was indistinguishable from the curve the United States.
Argentina was included to show a group with no identified DIF.
The two major standout conclusions from the plot are that India has non-uniform DIF with a flatter curve and that Indonesia has uniform DIF with the curve shifted to the left.
For India, this means that the item Happiness differentiates among people within the country with high happiness versus low happiness less well.
Relative to other countries, this means that, relative to the average across countries, individuals with higher happiness have a slightly lower observed response than we would expect to see from participants from other countries while individuals with lower happiness would have the opposite effect with an higher observed score relative to what we would expect if the item operated the same across countries. An alternative explanation, given the lack of differentiation between the item discrimination and latent response variable standard deviation, is that individuals in India are simply more varied in how they respond to items about happiness.

For Indonesia, we found that individuals are likelihood to have a higher observed response for all levels of the latent variable; though such differences are negligible for levels of the latent variable.
This implies that, for individuals with the same observed score, the part of the latent variable distribution being assessed tends to be individuals with lower happiness.
With only a single-item assessment, we cannot differentiate between this difference and a structural difference, meaning individuals in Indonesia may be happier OR may have a tendency to respond higher on such questions.

\section*{Concluding Remarks}

The present paper provides the first, to the best of my knowledge, approach for evaluating differential item function or measurement invariance of a single-item assessment \citep{Raudenska2023-bj, Raudenska2024-zh}.
Existing approaches have required using proxy items to create a ``pseudo-scale'' from items of a similar kind, but this requires additional leaps in logic that I was not comfortable with.
The proposed regularized HETOP model provides a way to evaluate DIF with only responses to the single-item and group membership.
The proposed method is certainly not recommended if a multi-item scale is available, but may be a path forward when not.
One major limitation is the inability to differentiate between DIF in the latent response variance (or scale) and the discrimination parameters--slope of the item characteristic curve. 
The variance and discrimination are inverses of each other in the probit model parameterization.
This lack of differentiation does not seem particularly important in item response theory applications \citep{De_Ayala2009}, so this may be an acceptable trade-off for the purposes of DIF evaluation.

\subsection*{Acknowledgments}
I would like to thank Drs. Tyler VanderWeele and Byron Johnson for their support throughout the Global Flourishing Study, which has supplied the impetus for the current work.


\bibliographystyle{apalike}
\bibliography{regHETOP}

\end{document}